\documentclass [12pt]{article}
\usepackage{graphicx,amssymb,amsmath,bm}
\usepackage[symbol*]{footmisc}
\usepackage[cp1251]{inputenc}
\usepackage[english]{babel}
\usepackage{booktabs}
\usepackage{cite}

\makeatletter
\def\@biblabel#1{#1.\hskip-0.3em}
\makeatother

\begin{document}
\def\refname{\normalsize \centering \mdseries \bf References}
\def\abstractname{Abstract}

\begin{center}
{\large \bf Polarized-deuteron scattering by spin-zero target nuclei at intermediate energies}
\end{center}

\begin{center}
{Valery~I.~Kovalchuk}
\footnote{E-mail: sabkiev@gmail.com}
\end{center}

\begin{center}
\small
\textit{Faculty of Physics, Taras Shevchenko National University of Kyiv,\\
64/13 Volodymyrs’ka Str., Kyiv 01601, Ukraine}
\end{center}

\begin{abstract}
General analytical expressions for the observables in $dA$-scattering reaction have been derived
in the diffraction approximation. The resulting formulas describe the cross section and polarization states
of the deuteron when it scattering by nuclei with zero spin in the ground state. The tabulated
distributions of the target nucleus density and the realistic deuteron wave functions calculated on the
basis of Nijmegen nucleon-nucleon potentials were used. The nucleon–nucleus phases were calculated in
the framework of Glauber formalism and making use of the double-folding potential. The calculated cross
sections and analyzing powers in elastic scattering of deuterons by $^{16}\text{O}$ and $^{40}\text{Ca}$
nuclei at 700~MeV are compared with the corresponding experimental data.
\vskip5mm
\flushleft
{\bf{Keywords:}} deuteron-nucleus scattering; eikonal approximation; double-folding potential
\end{abstract}

\bigskip
\begin{center}
\bf{1.~Introduction}
\end{center}
\smallskip

The scattering of deuterons by complex nuclei is of particular interest from the point of view of studying
the structure and interaction of the simplest composite nucleus with other nuclei. The peculiarity of
deuteron scattering from nuclei is determined by the features of its bound state. Since the deuteron spin
is equal to unity, the corresponding spin matrices are three-row, so their complete set can be represented
by five independent components~\cite{Sitenko2012}. Hence it follows that $dA$-scattering will be characterized
by a variety of polarization observables. The next feature of the deuteron is the non-sphericity of its
spatial shape, which manifests itself in the presence of a D-state to its wave function.
Finally, the deuteron is a weakly bound system that is reflected in the long tail of its density distribution,
and this feature clearly manifests itself, for example, in direct nuclear reactions involving deuterons
~\cite{Butler1957,Satchler1983}. Since polarization effects are mainly peripheral, the asymptotic behavior
of the deuteron wave function also becomes crucial for correct calculations.

The modern approach to describing collisions of light ions with atomic nuclei reflects the intention to create
a unified theory that includes both a realistic model of target nuclei and the microscopic optical potential
of nucleus-nucleus interaction~\cite{Pilipenko2015}. For this purpose, the folding model is most often used
(see, e.g.,~\cite{Satchler1979, Brandan1997, Lukyanov2007}), as well as its various variants developed within
the framework of the Glauber-type eikonal approximation (~\cite{Charagi1990, Hencken1996, Pinto1997, Lukyanov2015}
and references therein). In~\cite{Kovalchuk2015}, the formula for the eikonal phase was generalized for using of
realistic nuclear density distributions, that were expanded in the full Gaussoid basis. Together with the
Gaussian expansions of the wave functions under the integral sign in the expression for the reaction amplitude,
this made it possible to perform analytical integration and obtain general expressions for reaction observables
in the form of multiple sums with elementary functions. This approach is due to the fact that the general formulas
for the cross section and polarization in the problems of diffraction $dA$-scattering are quite inconvenient for
direct numerical calculations, therefore, for practical purposes, they are usually modified by introducing
additional simplifications and restrictions (e.g., the nucleus is opaque and non-diffuse; the deuteron radius
is much smaller than the target one; and so on). Notice that, in diffraction approximation, the reaction density
matrix is a five-fold integral only formally, because the profile functions, which the density matrix depends on,
are also expressed in terms of multiple integrals. Therefore, generally speaking, we have rather a complicated
computational problem. Nevertheless, the final formulas for the cross section and analyzing powers can be reduced
to algebraic expressions if the integrands are expanded into a series of the form
\begin{equation}
\Psi(x)=\sum_{j=1}^{N}a_{j}|\psi_{j}\rangle=
\sum_{j=1}^{N}a_{j}\exp{(-b_{j}x^2)}.
\label{eq1}
\end{equation}

Since the basis $|\psi_{j}\rangle$ is complete, any square-integrable function in some region can be expanded
in the same region with an arbitrary degree of accuracy. It should be noted that a similar approach is used
often enough: in the variational method for obtaining the energy levels of a bound system~\cite{Varga1995,
Kukulin1977, Grinyuk2014}, for the parametrization of nuclear charge densities in the ground state of
nucleus~\cite{DeVries1987, Sick1974}, in problems dealing with scattering~\cite{Dalkarov1985}, deuteron
stripping~\cite{Kovalchuk2015, Kovalchuk2016-1, Kovalchuk2018}, and fragmentation of light
nuclei~\cite{Kovalchuk2016-2, Kovalchuk2022}.

The structure of the paper is as follows. Section~2 is devoted to the description of formalism applied while
calculating the $dA$-scattering observables. In Sec.~3, the results of numerical calculations for the
differential cross sections and analyzing powers are discussed and compared with the corresponding experimental
data. Section~4 contains conclusions. The most important auxiliary formulas used to simplify the formalism
description are given in appendices.

\bigskip
\begin{center}
\bf{2.~Formalism}
\end{center}
\smallskip

All of the calculations that follow were made in the center-of-mass system using the system of units $\hbar=c=1$,
the Coulomb interaction was not taken into account. To describe the observables, we will use the formalism of
the density matrix. Let us expand the density matrix of the system $\rho$ in terms of the complete orthogonal
set of spin tensors $T_{IM}$~\cite{Sitenko2012}

\begin{equation}
\rho=\frac{1}{3}\sum_{I=0}^{2}\sum_{M=-I}^{I}\langle{T_{IM}^{\dagger}}\rangle T_{IM},
\label{eq2}
\end{equation}
where $T_{IM}$ values are expressed in terms of the components of the deuteron spin operator $\mathbf{S}$ in the
Cartesian coordinate system as follows~\cite{Lakin1955}
\begin{eqnarray}
T_{00}=1,\quad
T_{10}=\sqrt{\frac{3}{2}}\,S_{z},\quad
T_{11}=-\frac{\sqrt{3}}{2}\,(S_{x}+iS_{y}),\nonumber\\
T_{20}=\frac{1}{\sqrt{2}}\,(3S_{z}^{2}-2),\quad
T_{21}=-\frac{\sqrt{3}}{2}\,[(S_{x}+iS_{y})S_{z}+S_{z}(S_{x}+iS_{y})],\nonumber\\
T_{22}=\frac{\sqrt{3}}{2}\,(S_{x}+iS_{y})^{2},\quad
T_{i-j}=(-1)^{j}T_{ij}^{\dagger}.
\label{eq3}
\end{eqnarray}

The polarization state of the scattered deuteron is completely described by the average values of the spin tensors
\begin{equation}
\langle{T_{IM}}\rangle=\frac{\text{Tr}\,(\rho\,T_{IM})}{\text{Tr}\,{\rho}},
\label{eq4}
\end{equation}
where $\rho=F_{d}F_{d}^{\dagger}$ is the density matrix, $F_{d}$ is the deuteron-nucleus
scattering amplitude.

In the Cartesian coordinate system, the $z$-axis of which coincides with the quantization axis (field direction),
the polarization of the incident deuteron beam is described by two non-zero quantities~\cite{Nielsen2011}: $p_{z}$
and $p_{zz}$. Let $N_{0}$, $N_{+}$, $N_{-}$ are the numbers of deuterons with zero, up, and down spin projections,
respectively. Then the beam polarization asymmetry along the direction of the quantization axis is called the vector
polarization and is described as
\begin{equation}
p_{z}=\frac{N_{+}-N_{-}}{N_{+}+N_{-}+N_{0}}.
\label{eq5}
\end{equation}
The beam polarization asymmetry in the plane perpendicular to the quantization axis is called the tensor polarization
and is defined as follows
\begin{equation}
p_{zz}=\frac{N_{+}+N_{-}-2N_{0}}{N_{+}+N_{-}+N_{0}}.
\label{eq6}
\end{equation}
In a spherical coordinate system, the vector and tensor polarizations are related to the quantities (\ref{eq5}),
(\ref{eq6}) by the formulas
\begin{equation}
t_{10}=\sqrt{\frac{3}{2}}\,p_{z},\quad
t_{20}=\frac{1}{\sqrt{2}}\,p_{zz}.
\label{eq7}
\end{equation}

According to the Madison convention~\cite{Barschall1971}, measured polarization is defined in the right-handed
frame of references with the $z$-axis in the direction of the incident momentum $\mathbf{k}$ and the $y$-axis
in the direction of \mbox{$\mathbf{k}\times\mathbf{k}'$}, where $\mathbf{k}'$ is the momentum of the scattered
particles (Fig.~1).
%
\begin{figure}[!h]
\center
\includegraphics [scale=0.5] {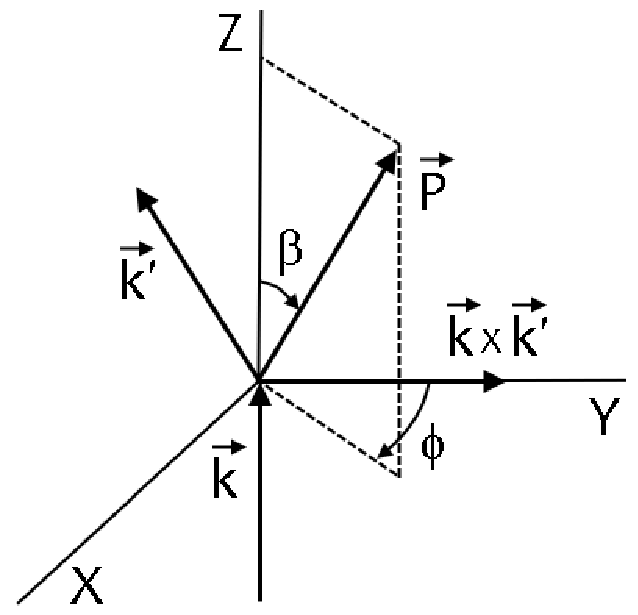}
\caption{The quantization axis $\mathbf{P}$ and the right-handed coordinate system as defined by the
Madison convention~\cite{Barschall1971}.}
\label{fig1}
\end{figure}

Transition to a new coordinate system whose quantization axis $\mathbf{P}$ is determined by the rotation
angles $\beta$ and $\phi$, somewhat complicates the expression for the cross section, which will also
depend on these angles. Using the polarization components (\ref{eq7}), the differential cross section
can be written as~\cite{Nielsen2011}
\begin{eqnarray}
\sigma(\theta,\phi)=\sigma_{0}(\theta)(1+G_{11}+G_{20}+G_{21}+G_{22}),\nonumber\\
G_{11}=\sqrt{2}\sin{\beta}\cos{\phi}\,\langle{iT_{11}}\rangle\,t_{10},\nonumber\\
G_{20}=(3\cos{\beta}^{2}-1)/2\,\langle{T_{20}}\rangle\,t_{20},\nonumber\\
G_{21}=\sqrt{3/2}\sin{2\beta}\cos{\phi}\,\langle{T_{21}}\rangle\,t_{20},\nonumber\\
G_{22}=-\sqrt{3/2}\sin{\beta}^{2}\cos{2\phi}\,\langle{T_{22}}\rangle\,t_{20},
\label{eq8}
\end{eqnarray}
where $\sigma_{0}(\theta)$ is the scattering cross section for unpolarized particles, and $\langle{iT_{11}}\rangle$,
$\langle{T_{20}}\rangle$, $\langle{T_{21}}\rangle$, $\langle{T_{22}}\rangle$ are quantities (\ref{eq4}).

Formulas (\ref{eq8}) are general. Usually, the experiment geometry is chosen in such a way that
$\beta=\pi/2$, $\phi=0$ (see, for example,~\cite{Arvieux1984,Nguyen1987}), then
\begin{equation}
\sigma(\theta)=\sigma_{0}(\theta)
[1+\sqrt{2}\,\langle{iT_{11}}\rangle\,t_{10}
-\frac{1}{2}\,\langle{T_{20}}\rangle\,t_{20}
-\sqrt{\frac{3}{2}}\,\langle{T_{22}}\rangle\,t_{20}].
\label{eq9}
\end{equation}
Introducing
\begin{equation}
A_{y}=\frac{2}{\sqrt{3}}\langle{iT_{11}}\rangle,\quad
A_{yy}=-\frac{1}{\sqrt{2}}\langle{T_{20}}\rangle-\sqrt{3}\langle{T_{22}}\rangle,
\label{eq10}
\end{equation}
where \mbox{$A_{y}=A_{y}(\theta)$}, \mbox{$A_{yy}=A_{yy}(\theta)$} are the polarizations (the analyzing powers)
of outgoing particles, expression (\ref{eq9}) can be written as
\begin{equation}
\sigma(\theta)=\sigma_{0}(\theta)
[1+\sqrt{\frac{3}{2}}\,A_{y}(\theta)\,t_{10}+\frac{1}{\sqrt{2}}\,A_{yy}(\theta)\,t_{20}].
\label{eq11}
\end{equation}

Thus, the scattering cross section depends on the analyzing powers of the reaction and on the polarizations
of the incident beam, which are specified by the experimental conditions.

In the Glauber approximation, the amplitude of $dA$-scattering defined as~\cite{Sitenko1990}
\begin{equation}
F_{d}(\mathbf{q})=\frac{ik}{2\pi}\int d\mathbf{b}\exp(i{\bf qb})\int d{\mathbf{r}}\,
\varphi_{0}^{\dagger}(\mathbf{r}) (\Omega_{1}+\Omega_{2}-\Omega_{1}\Omega_{2})
\varphi_{0}(\mathbf{r}),
\label{eq12}
\end{equation}
where $\mathbf{q}$ is the momentum transfer, $\mathbf{b}$ is the impact parameter of deuteron center of mass,
$\varphi_{0}$ is the wave function of deuteron ground state, $\Omega_{n}$ are the neutron-nucleus $(n=1)$ and
proton-nucleus $(n=2)$ profile functions, which are operators
\begin{equation}
{\Omega}_{n}(\mathbf{r}_{n}) = \omega_{n}(b_{n})\{1 + \gamma_{n}\exp(i\delta_{n})
{\bm\sigma}_{n}((\mathbf{k}/2)\!\times\!\nabla_{n})\}.
\label{eq13}
\end{equation}
The quantities $\gamma_{n}$ and $\delta_{n}$ in (\ref{eq13}) are the parameters of the spin-orbit interaction,
${\bm\sigma}_{n}$ are the Pauli matrices, $\nabla_{n}\equiv\partial/\partial\mathbf{r}_{n}$.

As $\varphi_{0}(\mathbf{r})$ in (\ref{eq12}), we use the deuteron wave function~\cite{Sitenko2012}
\begin{equation}
\varphi_{0}(\mathbf{r})=\varphi_{\text{S}}(r)+\varphi_{\text{D}}(r)
\mathbf{S}_{12}(\mathbf{r},\mathbf{S}),
\label{eq14}
\end{equation}
where $\varphi_{\text{S}}$, $\varphi_{\text{D}}$ are the radial components that describe
the S- and D-states of the deuteron, \mbox{$\mathbf{S}_{12}=6(\mathbf{S}\mathbf{r})^2/r^2-2{\mathbf{S}}^2$}
is the nucleon-nucleon spin operator, \mbox{$\mathbf{S}=({\bm\sigma}_{1}+{\bm\sigma}_{2})/2$} is the
deuteron total spin.

The amplitude (\ref{eq12}) and the average values of the spin-tensor components (\ref{eq4}) can be calculated
using the microscopic approach described in~\cite{Kovalchuk2015}. As the radial components of $\varphi_{0}(\mathbf{r})$,
we use their tabulated values for Nijmegen potentials~\cite{Stoks1994}, which we expand in series of Gaussoid
basis functions
\begin{equation}
\varphi_{0}(\mathbf{r})=\sum_{j=1}^{N}g_{j}\exp(-\lambda_{j}r^2)+
h_{j}r^{2}\exp(-\mu_{j}r^2)\mathbf{S}_{12}(\mathbf{r},\mathbf{S}).
\label{eq15}
\end{equation}
We calculate $\omega_{n}$ in (\ref{eq13}) using the eikonal approximation (see Appendix A) and
also expand them in the same way as in (\ref{eq15}):
\begin{equation}
\omega_{n}(b_{n})=\sum_{j=1}^{N}\alpha_{nj}\exp(-b_{j}^{2}/d_{nj}),\quad
d_{nj}=R_{\text{rms}}^{2}/j,
\label{eq16}
\end{equation}
where $R_{\text{rms}}$ is the root-mean-square radius of the target nucleus.

The integral (\ref{eq12}) with the functions (\ref{eq15}), (\ref{eq16}) is calculated analytically, so that
the final result (the cross section $\sigma_{0}$ and the values of $\langle{T_{IM}}\rangle$) can be
presented as multiple sums.

Now, substituting (\ref{eq13}), (\ref{eq15}), (\ref{eq16}) into (\ref{eq12}), we obtain (see also~\cite{Sitenko2012}):
\begin{equation}
\sigma_{0}=\frac{1}{3}\,\text{Tr}(F_{d}F_{d}^{\dagger})=
|A|^2+\frac{2}{3}\,|B|^2+\frac{2}{9}\,|C|^2+\frac{2}{3}\,|D|^2,
\label{eq17}
\end{equation}
\begin{equation}
\langle{T_{10}}\rangle=\frac{1}{3}\,\text{Tr}(F_{d}F_{d}^{\dagger}\,{T_{10}})=0,
\label{eq18}
\end{equation}
\begin{equation}
\langle{iT_{11}}\rangle=\frac{1}{3}\,\text{Tr}(F_{d}F_{d}^{\dagger}\,{iT_{11}})=
\frac{\sqrt{2}}{3\sigma_{0}}\,\text{Re}\Bigl(A-\frac{1}{3}\,C\Bigr)B^{*},
\label{eq19}
\end{equation}
\begin{equation*}
\langle{T_{20}}\rangle=\frac{1}{3}\,\text{Tr}(F_{d}F_{d}^{\dagger}\,{T_{20}})=
\end{equation*}
\begin{equation}
=\frac{\sqrt{2}}{3\sigma_{0}}\,\Bigl\{\text{Re}\,A(C-3D)^{*}+\text{Re}\,CD^{*}-
\frac{1}{2}\,|B|^{2}-\frac{1}{6}\,|C|^{2}+\frac{1}{2}\,|D|^{2} \Bigr\},
\label{eq20}
\end{equation}
\begin{equation}
\langle{T_{21}}\rangle=\frac{1}{3}\,\text{Tr}(F_{d}F_{d}^{\dagger}\,{T_{21}})=
\frac{\sqrt{2}}{3\sigma_{0}}\,\text{Im}\,BD^{*},
\label{eq21}
\end{equation}
\begin{equation*}
\langle{T_{22}}\rangle=\frac{1}{3}\,\text{Tr}(F_{d}F_{d}^{\dagger}\,{T_{22}})=
\end{equation*}
\begin{equation}
=\frac{1}{\sqrt{3}\sigma_{0}}\,\Bigl\{\text{Re}\,A(C+D)^{*}-\frac{1}{3}\,\text{Re}\,CD^{*}-
\frac{1}{2}\,|B|^{2}+\frac{1}{6}\,|C|^{2}-\frac{1}{2}\,|D|^{2}\Bigr\}.
\label{eq22}
\end{equation}

\vspace{15pt}
\noindent
The values of $A$, $B$, $C$, and $D$ are the result of integration in (\ref{eq12}) and are defined as follows:
\begin{equation*}
A=\frac{ik\pi^{3/2}}{2}\,\Bigl\{A_{1}-4A_{2}+2A_{3}-
64A_{4}+3q^{2}A_{5}-
\end{equation*}
\begin{equation}
-384[\gamma_{1}\exp(i\delta_{1})+\gamma_{2}\exp(i\delta_{2})]A_{6}\Bigr\},
\label{eq23}
\end{equation}
\begin{equation*}
B=\frac{qk^{2}\pi^{3/2}}{4}\,\Bigl\{B_{1}-B_{2}+\sqrt{2}q^{2}B_{3}-
\end{equation*}
\begin{equation}
-2[\gamma_{1}\exp(i\delta_{1})+\gamma_{2}\exp(i\delta_{2})]\bigl(
B_{4}-8B_{5}-\sqrt{8}B_{6}\bigr)\Bigr\},
\label{eq24}
\end{equation}
\begin{equation}
C=\frac{3iqk^{2}\pi^{3/2}}{16}\,\Bigl\{C_{1}-12C_{2}
-16\sqrt{2}\bigl(C_{3}-6C_{4}\bigr)\Bigr\},
\label{eq25}
\end{equation}
\begin{equation*}
D=12ik\pi^{3/2}\,\Bigl\{D_{1}-\sqrt{8}D_{2}+
\end{equation*}
\begin{equation}
+[\gamma_{1}\exp(i\delta_{1})+\gamma_{2}\exp(i\delta_{2})]\bigl(
6D_{3}+3\sqrt{8}D_{4}\bigr)\Bigr\}.
\label{eq26}
\end{equation}

Thus, the complete set of observables (\ref{eq17})-(\ref{eq22}) is defined by
20 functions in the expressions (\ref{eq23})-(\ref{eq26}) (see Appendix B).

\bigskip
\begin{center}
\bf{3.~Calculation results and discussion}
\end{center}
\smallskip

The formalism described in the previous section was applied to analyze experimental data
obtained for the polarized deuteron scattering from the $^{16}\text{O}$ and $^{40}\text{Ca}$
nuclei at projectile energy of 700~MeV~\cite{Nguyen1987, Nguyen1985}. Description of the
experimental setup is given in~\cite{Arvieux1984}, according to which the polarizations
(\ref{eq7}) were defined as
\begin{equation}
t_{10}=\frac{1}{\sqrt{6}}\,P_{V},\quad
t_{20}=\frac{1}{\sqrt{2}}\,P_{T},
\label{eq27}
\end{equation}
where $P_{V}$, $P_{T}$ are the degrees of vector and tensor polarization of a polarized deuteron beam,
the numerical values of which are given in~\cite{Nguyen1987, Nguyen1985}.

The spin-orbit interaction parameters in (\ref{eq13}) were approximately determined from the
experiments~\cite{Kelly1991-1, Kelly1991-2, Frekers1987} on measuring the polarization of protons
acquired by them upon scattering from $^{16}\text{O}$ and $^{40}\text{Ca}$ nuclei.
As is known~\cite{Fermi1954}, the angular dependence of the nucleon polarization $P(\theta)$ at
intermediate energies is described by the Fermi formula, from which it follows that
\mbox{$\delta=\text{arcsin}\,P(\theta_{\text{max}})$}, where $\theta_{\text{max}}$ is the scattering angle
of maximum polarization; \mbox{$\gamma=(kq_{\text{max}})^{-1}$}, where $k$ is the nucleon momentum,
$q_{\text{max}}$ is the momentum transfer at \mbox{$\theta=\theta_{\text{max}}$}.
From the data of work~\cite{Kelly1991-1, Kelly1991-2, Frekers1987}, it follows that the parameters
\mbox{$\delta=\delta_{1,2}$} and \mbox{$\gamma=\gamma_{1,2}$} are equal to 0.77 and 0.31$\text{fm}^{2}$,
respectively, for the $^{16}\text{O}$ target nucleus, and to 0.64 and 0.34$\text{fm}^{2}$, respectively,
for $^{40}\text{Ca}$.

Figure 2 shows the calculation results of the observables in the reaction of polarized deuteron scattering
from $^{16}\text{O}$ and $^{40}\text{Ca}$ nuclei at 700~MeV.

\begin{figure}[!h]
\center
\includegraphics [scale=1.0] {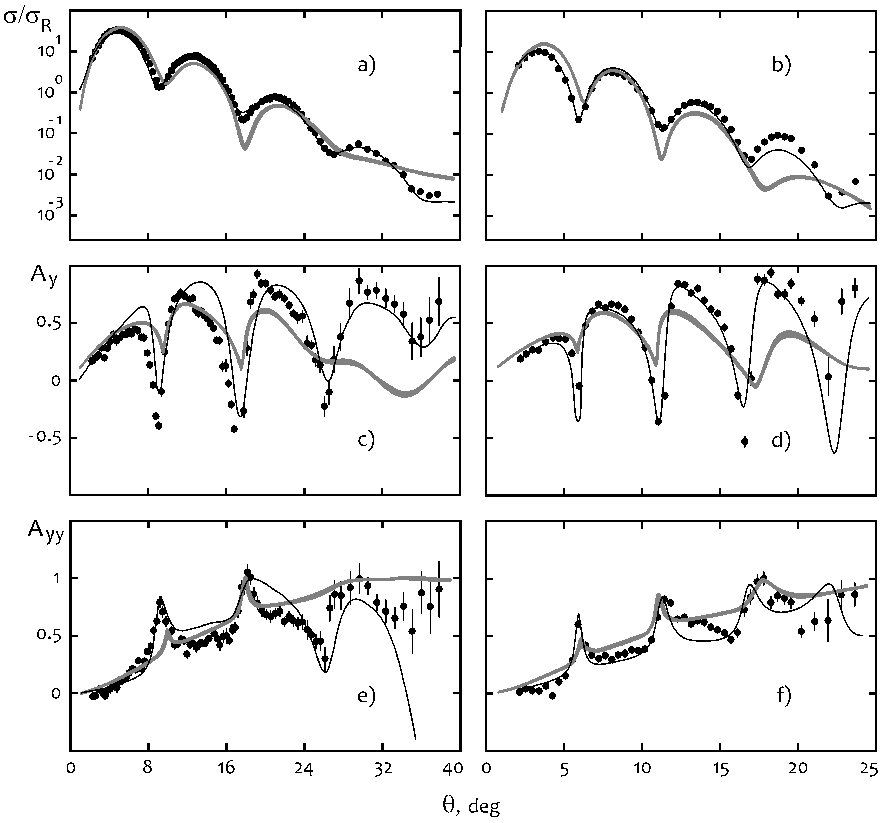}
\caption{Cross sections \mbox{$\sigma=\sigma(\theta)$}, relative to Rutherford one
\mbox{$\sigma_{R}=\sigma_{R}(\theta)$}, and analyzing powers \mbox{$A_{y}(\theta)$},
\mbox{$A_{yy}(\theta)$} in scattering reaction of polarized deuterons by \mbox{$^{16}\text{O}$} (a,c,e)
and \mbox{$^{40}\text{Ca}$} (b,d,f) nuclei at 700~MeV. Experimental data were taken
from~\cite{Nguyen1987,Nguyen1985}. See text for details.}
\label{fig2}
\end{figure}

When expanding the wave function $\varphi_{0}(\mathbf{r})$ in series (\ref{eq15}), tabulated data~\cite{Stoks1994}
for the S- and D-component of deuteron wave function obtained with the help of realistic $NN$-potentials
Nijm~I, Nijm~II, Nijm~93, and Reid~93 were used. The curves corresponding to these potentials lie inside
the dark gray bands (see Fig.~2). The profile functions $\omega_{n}(b_{n})$ in (\ref{eq13}), before a series
expansion (\ref{eq16}), were first calculated in the eikonal approximation making use of the tabulated
nuclear density distributions for the $^{16}\text{O}$ and $^{40}\text{Ca}$ nuclei taken from~\cite{DeVries1987}.
The number of expansion terms in (\ref{eq15}), (\ref{eq16}) was $N=12$.

In preliminary calculations, it was found that the strong absorption model with one parameter $N_{W}$ (see Appendix A)
is able to describe the observables~\cite{Nguyen1987,Nguyen1985} only for a small range of scattering angles
\mbox{$\theta<10^{\circ}$} (these results are not shown in Fig.~2). Therefore, to improve the description of
experiments, a semi-microscopic approach~\cite{Satchler1979} was used. The real part of the $NN$-potential was
added to the expression for the eikonal phase (\ref{A2}) in the form
\begin{equation}
V(r)=V_{0}\Bigl[1+\exp\Bigl(\frac{r-R_{V}}{a_{V}}\Bigr)\Bigr]^{-1}.
\label{eq28}
\end{equation}
The depth of potential (\ref{eq28}) was given by the relation \mbox{$V_{0}=\bar\alpha_{NN}W_{0}$},
where $\bar\alpha_{NN}$ is the nuclear isospin-averaged ratio of the real to the imaginary part of the amplitude
for nucleon–nucleon scattering at zero angle. The values of $\bar\alpha_{NN}$ were taken from~\cite{Shukla2001},
the numerical values of \mbox{$(r_{V}; a_{V})$} were equal to \mbox{$(0.98; 0.15)$} for the $^{16}\text{O}$
target nucleus, and \mbox{$(1.08; 0.15)$} for $^{40}\text{Ca}$.

From the analysis of the behavior of the calculated curves in Fig.~2, it follows that for both target nuclei:

i) experimental differential cross sections are satisfactorily described in the first three maxima
and minima;

ii) experimental values of $A_{y}$ are satisfactorily described in the first two maxima;

iii) experimental values of $A_{yy}$ are described both qualitatively and quantitatively.

For comparison, in Fig.~2 shows the results of works by other authors, where
experiments~\cite{Nguyen1987,Nguyen1985} were also analyzed (solid black curves). In particular,
to describe the scattering of deuterons by $^{16}\text{O}$ nuclei~\cite{Nguyen1987},
an optical model with nine parameters was used, and for $^{40}\text{Ca}$, a relativistic folding model
with a phenomenological parametrization of the $NN$-interaction~\cite{Pinto1996} was used.

From the analysis of the calculation results presented in Fig.~2, it follows that:

i) the optical model~\cite{Nguyen1987} provides the best description of the experiment;
in our opinion, this is achieved by increasing the number of parameters of the optical model;

ii) the semi-microscopic formalism presented in this paper and Dirac's microscopic
formalism~\cite{Pinto1996} lead to qualitatively identical results; it should be noted that
agreement with experiment in~\cite{Pinto1996} improved significantly only when the multiple
scattering effects were taking into account.

\bigskip
\begin{center}
\bf{4.~Conclusions}
\end{center}
\smallskip

The paper describes spin-dependent observables in the reaction of polarized deuteron scattering by spin-zero
target nuclei at intermediate energies. Within the framework of the Glauber model, general analytical expressions
are obtained for the reaction cross section and average values of the spin-tensor components.

The calculations used a semi-microscopic approach, in which neither the deuteron wave function nor
the nucleon-nucleus profile functions were modeled. This made it possible to minimize the set of fitting
parameters and establish, firstly, that:

i) the diffraction model of strong absorption makes allows to describe the experimental data only in
small range of scattering angles \mbox{$\theta<{10}^{\circ}$};

ii) the diffraction model with refraction (semi-microscopic approach) slightly expands the model applicability
up to \mbox{$\theta<{{{15}^{\circ}}\div{{20}^{\circ}}}$}, where the cross sections and tensor analysing powers
can be described; on the other hand, the vector analysing powers were described qualitatively, except the
region \mbox{$\theta<{10}^{\circ}$}; agreement with these experiments could be improved by using model profile
functions with additional parameters: this is usually done in an optical model, where the spin-orbit part of
the potential differs from the central one.

Secondly, the diffraction model itself and the limits of its applicability to the kinematics of the reactions
concerned were verified. The validity of the formula (\ref{eq12}) is determined by the conditions \mbox{$q\ll{k}$}
(or \mbox{$kR\gg1$}, where $R$ is the radius of the deuteron-nucleus interaction). If we take for $R$, for example,
the root-mean-square radius of the $^{16}\text{O}$ nucleus, which is equal to 2.711~fm~\cite{DeVries1987}, then for
deuteron energy of 700~MeV the inequality \mbox{$kR\gg1$} undoubtedly satisfied. The inequality \mbox{$q\ll{k}$} gives
the estimate \mbox{$\theta<\theta_{\text{max}}\simeq{10}^{\circ}$}, which is confirmed by calculations for
the diffraction model of strong absorption. As follows from the results obtained above, the refraction model
extends the applicability domain of (\ref{eq12}) up to \mbox{$\theta_{\text{max}}\simeq{20}^{\circ}$}.

\setcounter{section}{0}
\def\theequation{\Alph{section}.\arabic{equation}}
\def\thesection{\normalsize Appendix \Alph{section}}

\setcounter{equation}{0}
\section{}
\hspace{\parindent}
The radial parts of nucleon-nucleus profile functions were calculated in the eikonal approximation:
\begin{equation}
\omega_{i}(b_{i})=1-\exp[-\phi_{i}(b_{i})],\quad i\!=\!1,2;
\label{A1}
\end{equation}
where
\begin{equation}
\phi_{i}(b_{i})=-\frac{1}{v}\int_{-\infty}^{\infty}dz\,W\left(\sqrt{b_{i}^{2}+z^{2}}\right)
\label{A2}
\end{equation}
is the scattering phase, $v$ the velocity of incident nucleon, and $W(r)$ the imaginary part of
nucleon-nucleus potential.

In the framework of the double folding model, the eikonal phase can be calculated using the method
described in work~\cite{Charagi1990}. Let the distribution of nuclear density in the nucleon,
$\rho_{i}(r)$, and the amplitude of $NN$-interaction at the impact parameter plane, $f(b)$,
be defined by Gaussian functions:
\begin{equation}
\rho_{i}(r)=\rho_{i}(0)\exp(-r^{2}/a_{i}^{2}),
\label{A3}
\end{equation}
\begin{equation}
f(b)=(\pi r_{0}^{2})^{-1}\exp(-b^{2}/r_{0}^{2}),
\label{A4}
\end{equation}
where $\rho_{i}(0)=(a_{i}\sqrt{\pi})^{-3}$, $a_{i}^{2}=r_{0}^{2}=2r_{NN}^{2}/3$,
and $r_{NN}^{2}\cong~0.65~\mathrm{fm}^{2}$ is the mean-square radius of $NN$-interaction. If the
density distribution (tabulated~\cite{DeVries1987} or model) in the target nucleus can be expanded in
a series of Gaussoid basis functions,
\begin{equation}
\rho_{T}(r)=\sum_{j=1}^{N}\rho_{Tj}\exp(-r^{2}/a_{Tj}^{2}),\quad
a_{Tj}^{2}=R_{rms}^{2}/j\,,
\label{A5}
\end{equation}
where $R_{rms}$ is the root-mean-square radius of the nucleus, the formula for the eikonal phase
from work~\cite{Charagi1990} can be generalized~\cite{Kovalchuk2015} to the expression
\begin{equation}
\phi_{i}(b_{i})=N_{W}\sqrt{\pi}\,\bar{\sigma}_{NN}\sum_{j=1}^{N}
\frac{\rho_{Tj}\,a_{Tj}^{3}}{a_{Tj}^{2}+2r_{0}^{2}}
\exp\left(-\frac{b_{i}^{2}}{a_{Tj}^{2}+2r_{0}^{2}}\right),
\label{A6}
\end{equation}
where $N_{W}$ is the normalization factor for the imaginary part of the double folding potential,
and $\bar{\sigma}_{NN}$ is the isotopically averaged cross-section of nucleon-nucleon
interaction~\cite{Shukla2001}.

Formula (\ref{A6}) was used directly while calculating profile functions (\ref{A1}). Afterwards,
they were expanded in the Gaussoid basis (see (\ref{eq16})).

\setcounter{equation}{0}
\section{}
\hspace{\parindent}
The quantities included in formulas (\ref{eq23})-(\ref{eq26}) are obtained as a result of
analytical integration in (\ref{eq12}) and are determined as follows:
\begin{equation*}
A_{1}=\sum_{j=1}^{N}\sum_{\ell=1}^{N}\sum_{m=1}^{N}
\Bigl(
\alpha_{1j}\beta_{1j}\exp\Bigl(-\frac{\beta_{1j}q^2}{4}\Bigr)+
\alpha_{2j}\beta_{2j}\exp\Bigl(-\frac{\beta_{2j}q^2}{4}\Bigr)
\Bigr)\times
\end{equation*}
\begin{equation}
\times\frac{g_{\ell}g_{m}}{\lambda_{\ell m}^{3/2}}
\exp\Bigl(-\frac{q^2}{16\lambda_{\ell m}}\Bigr),
\label{B1}
\end{equation}

\begin{equation}
A_{2}=\sum_{i=1}^{N}\sum_{j=1}^{N}\sum_{\ell=1}^{N}\sum_{m=1}^{N}
\frac{\alpha_{1j}\alpha_{2j}\beta_{ij}g_{\ell}g_{m}}
{\bigl(\beta_{ij}^{-1}+4\lambda_{\ell m}\bigr)\lambda_{\ell m}^{1/2}}
\exp\Bigl(-\frac{\beta_{ij}q^2}{4}\Bigr),
\label{B2}
\end{equation}

\begin{equation*}
A_{3}=\sum_{j=1}^{N}\sum_{\ell=1}^{N}\sum_{m=1}^{N}
\Bigl(
\alpha_{1j}\beta_{1j}\exp\Bigl(-\frac{\beta_{1j}q^2}{4}\Bigr)+
\alpha_{2j}\beta_{2j}\exp\Bigl(-\frac{\beta_{2j}q^2}{4}\Bigr)
\Bigr)\times
\end{equation*}
\begin{equation}
\times\frac{h_{\ell}h_{m}}{\mu_{\ell m}^{7/2}}
\Bigl(15-\frac{5q^{2}}{4\mu_{\ell m}}+\frac{q^{4}}{64\mu_{\ell m}^{2}}\Bigr)
\exp\Bigl(-\frac{q^2}{16\mu_{\ell m}}\Bigr),
\label{B3}
\end{equation}

\begin{equation}
A_{4}=\sum_{i=1}^{N}\sum_{j=1}^{N}\sum_{\ell=1}^{N}\sum_{m=1}^{N}
\frac{\alpha_{1j}\alpha_{2j}\beta_{ij}h_{\ell}h_{m}}
{\bigl(\beta_{ij}^{-1}+4\mu_{\ell m}\bigr)^{3}\mu_{\ell m}^{5/2}}
\Bigl(30\mu_{\ell m}^{2}+\frac{5\mu_{\ell m}}{\beta_{ij}}+\frac{3}{8\beta_{ij}^{2}}\Bigr)
\exp\Bigl(-\frac{\beta_{ij}q^2}{4}\Bigr),
\label{B4}
\end{equation}

\begin{equation*}
A_{5}=\sum_{j=1}^{N}\sum_{\ell=1}^{N}\sum_{m=1}^{N}
\Bigl(
\gamma_{1}\alpha_{1j}\beta_{1j}\exp\Bigl(-\frac{\beta_{1j}q^2}{4}+i\delta_{1}\Bigr)-
\gamma_{2}\alpha_{2j}\beta_{2j}\exp\Bigl(-\frac{\beta_{2j}q^2}{4}+i\delta_{2}\Bigr)
\Bigr)\times
\end{equation*}
\begin{equation}
\times\frac{h_{\ell}h_{m}}{\mu_{\ell m}^{7/2}}
\Bigl(5-\frac{q^{2}}{8\mu_{\ell m}}\Bigr)
\exp\Bigl(-\frac{q^2}{16\mu_{\ell m}}\Bigr),
\label{B5}
\end{equation}

\begin{equation}
A_{6}=\sum_{i=1}^{N}\sum_{j=1}^{N}\sum_{\ell=1}^{N}\sum_{m=1}^{N}
\frac{\alpha_{1j}\alpha_{2j}h_{\ell}h_{m}}
{\bigl(\beta_{ij}^{-1}+4\mu_{\ell m}\bigr)^{3}\mu_{\ell m}^{3/2}}
\Bigl(3\mu_{\ell m}-\frac{1}{4\beta_{ij}}\Bigr)
\exp\Bigl(-\frac{\beta_{ij}q^2}{4}\Bigr),
\label{B6}
\end{equation}

\begin{equation*}
B_{1}=\sum_{j=1}^{N}\sum_{\ell=1}^{N}\sum_{m=1}^{N}
\Bigl(
\gamma_{1}\alpha_{1j}\beta_{1j}\exp\Bigl(-\frac{\beta_{1j}q^2}{4}+i\delta_{1}\Bigr)+
\gamma_{2}\alpha_{2j}\beta_{2j}\exp\Bigl(-\frac{\beta_{2j}q^2}{4}+i\delta_{2}\Bigr)
\Bigr)\times
\end{equation*}
\begin{equation}
\times\frac{g_{\ell}g_{m}}{\lambda_{\ell m}^{3/2}}
\exp\Bigl(-\frac{q^2}{16\lambda_{\ell m}}\Bigr),
\label{B7}
\end{equation}

\begin{equation*}
B_{2}=\sum_{j=1}^{N}\sum_{\ell=1}^{N}\sum_{m=1}^{N}
\Bigl(
\gamma_{1}\alpha_{1j}\beta_{1j}\exp\Bigl(-\frac{\beta_{1j}q^2}{4}+i\delta_{1}\Bigr)+
\gamma_{2}\alpha_{2j}\beta_{2j}\exp\Bigl(-\frac{\beta_{2j}q^2}{4}+i\delta_{2}\Bigr)
\Bigr)\times
\end{equation*}
\begin{equation}
\times\frac{h_{\ell}h_{m}}{\mu_{\ell m}^{7/2}}
\Bigl(15-\frac{17q^{2}}{8\mu_{\ell m}}+\frac{q^{4}}{32\mu_{\ell m}^{2}}\Bigr)
\exp\Bigl(-\frac{q^2}{16\mu_{\ell m}}\Bigr),
\label{B8}
\end{equation}

\begin{equation*}
B_{3}=\sum_{j=1}^{N}\sum_{\ell=1}^{N}\sum_{m=1}^{N}
\Bigl(
\gamma_{1}\alpha_{1j}\beta_{1j}\exp\Bigl(-\frac{\beta_{1j}q^2}{4}+i\delta_{1}\Bigr)+
\gamma_{2}\alpha_{2j}\beta_{2j}\exp\Bigl(-\frac{\beta_{2j}q^2}{4}+i\delta_{2}\Bigr)
\Bigr)\times
\end{equation*}
\begin{equation}
\times\frac{g_{\ell}h_{m}}{(\lambda_{\ell}+\mu_{m})^{7/2}}
\exp\Bigl(-\frac{q^2}{8(\lambda_{\ell}+\mu_{m})}\Bigr),
\label{B9}
\end{equation}

\begin{equation}
B_{4}=A_{2},
\label{B10}
\end{equation}

\begin{equation}
B_{5}=\sum_{i=1}^{N}\sum_{j=1}^{N}\sum_{\ell=1}^{N}\sum_{m=1}^{N}
\frac{\alpha_{1j}\alpha_{2j}\beta_{ij}h_{\ell}h_{m}}
{\bigl(\beta_{ij}^{-1}+4\mu_{\ell m}\bigr)^{3}\mu_{\ell m}^{5/2}}
\Bigl(15\mu_{\ell m}^{2}+\frac{17\mu_{\ell m}}{4\beta_{ij}}+\frac{3}{8\beta_{ij}^{2}}\Bigr)
\exp\Bigl(-\frac{\beta_{ij}q^2}{4}\Bigr),
\label{B11}
\end{equation}

\begin{equation}
B_{6}=\sum_{i=1}^{N}\sum_{j=1}^{N}\sum_{\ell=1}^{N}\sum_{m=1}^{N}
\frac{\alpha_{1j}\alpha_{2j}g_{\ell}h_{m}}
{\bigl(\beta_{ij}^{-1}+2(\lambda_{\ell}+\mu_{m})\bigr)^{2}(\lambda_{\ell}+\mu_{m})^{3/2}}
\exp\Bigl(-\frac{\beta_{ij}q^2}{4}\Bigr),
\label{B12}
\end{equation}

\begin{equation*}
C_{1}=\sum_{j=1}^{N}\sum_{\ell=1}^{N}\sum_{m=1}^{N}
\Bigl(
\alpha_{1j}\beta_{1j}\exp\Bigl(-\frac{\beta_{1j}q^2}{4}\Bigr)+
\alpha_{2j}\beta_{2j}\exp\Bigl(-\frac{\beta_{2j}q^2}{4}\Bigr)
\Bigr)\times
\end{equation*}
\begin{equation}
\times\frac{h_{\ell}h_{m}}{\mu_{\ell m}^{9/2}}
\Bigl(7-\frac{q^{2}}{8\mu_{\ell m}}\Bigr)
\exp\Bigl(-\frac{q^2}{16\mu_{\ell m}}\Bigr),
\label{B13}
\end{equation}

\begin{equation*}
C_{2}=\sum_{j=1}^{N}\sum_{\ell=1}^{N}\sum_{m=1}^{N}
\Bigl(
\gamma_{1}\alpha_{1j}\beta_{1j}\exp\Bigl(-\frac{\beta_{1j}q^2}{4}+i\delta_{1}\Bigr)-
\gamma_{2}\alpha_{2j}\beta_{2j}\exp\Bigl(-\frac{\beta_{2j}q^2}{4}+i\delta_{2}\Bigr)
\Bigr)\times
\end{equation*}
\begin{equation}
\times\frac{h_{\ell}h_{m}}{\mu_{\ell m}^{7/2}}
\Bigl(1-\frac{q^{2}}{8\mu_{\ell m}}\Bigr)
\exp\Bigl(-\frac{q^2}{16\mu_{\ell m}}\Bigr),
\label{B14}
\end{equation}

\begin{equation*}
C_{3}=\sum_{j=1}^{N}\sum_{\ell=1}^{N}\sum_{m=1}^{N}
\Bigl(
\alpha_{1j}\beta_{1j}\exp\Bigl(-\frac{\beta_{1j}q^2}{4})+
\alpha_{2j}\beta_{2j}\exp\Bigl(-\frac{\beta_{2j}q^2}{4}\Bigr)
\Bigr)\times
\end{equation*}
\begin{equation}
\times\frac{g_{\ell}h_{m}}{(\lambda_{\ell}+\mu_{m})^{7/2}}
\exp\Bigl(-\frac{q^2}{8(\lambda_{\ell}+\mu_{m})}\Bigr),
\label{B15}
\end{equation}

\begin{equation*}
C_{4}=\sum_{j=1}^{N}\sum_{\ell=1}^{N}\sum_{m=1}^{N}
\Bigl(
\gamma_{1}\alpha_{1j}\beta_{1j}\exp\Bigl(-\frac{\beta_{1j}q^2}{4}+i\delta_{1}\Bigr)-
\gamma_{2}\alpha_{2j}\beta_{2j}\exp\Bigl(-\frac{\beta_{2j}q^2}{4}+i\delta_{2}\Bigr)
\Bigr)\times
\end{equation*}
\begin{equation}
\times\frac{g_{\ell}h_{m}\lambda_{\ell}}{4(\lambda_{\ell}+\mu_{m})^{7/2}}
\exp\Bigl(-\frac{q^2}{8(\lambda_{\ell}+\mu_{m})}\Bigr),
\label{B16}
\end{equation}

\begin{equation}
D_{1}=\sum_{i=1}^{N}\sum_{j=1}^{N}\sum_{\ell=1}^{N}\sum_{m=1}^{N}
\frac{\alpha_{1j}\alpha_{2j}h_{\ell}h_{m}}
{\bigl(\beta_{ij}^{-1}+4\mu_{\ell m}\bigr)^{3}\mu_{\ell m}^{5/2}}
\Bigl(\frac{3}{2\beta_{ij}}+14\mu_{\ell m}\Bigr)
\exp\Bigl(-\frac{\beta_{ij}q^2}{4}\Bigr),
\label{B17}
\end{equation}

\begin{equation}
D_{2}=B_{6},
\label{B18}
\end{equation}

\begin{equation}
D_{3}=\sum_{i=1}^{N}\sum_{j=1}^{N}\sum_{\ell=1}^{N}\sum_{m=1}^{N}
\frac{\alpha_{1j}\alpha_{2j}h_{\ell}h_{m}}
{\bigl(\beta_{ij}^{-1}+4\mu_{\ell m}\bigr)^{3}\mu_{\ell m}^{3/2}}
\Bigl(\frac{3}{2\beta_{ij}}-2\mu_{\ell m}\Bigr)
\exp\Bigl(-\frac{\beta_{ij}q^2}{4}\Bigr),
\label{B19}
\end{equation}

\begin{equation}
D_{4}=\sum_{i=1}^{N}\sum_{j=1}^{N}\sum_{\ell=1}^{N}\sum_{m=1}^{N}
\frac{\alpha_{1j}\alpha_{2j}g_{\ell}h_{m}\lambda_{\ell}}
{\bigl(\beta_{ij}^{-1}+2(\lambda_{\ell}+\mu_{m})\bigr)^{2}(\lambda_{\ell}+\mu_{m})^{3/2}}
\exp\Bigl(-\frac{\beta_{ij}q^2}{4}\Bigr).
\label{B20}
\end{equation}

Coefficients with double summation indices that appear in the formulas (\ref{B1})-(\ref{B20}),
are defined in terms of expansion coefficients (\ref{eq15}), (\ref{eq16}) as
\mbox{$\beta_{ij}=d_{1j}d_{2j}/(d_{1j}+d_{2j})$},
\mbox{$\lambda_{{\ell}m}=(\lambda_{\ell}+\lambda_{m})/2$},
\mbox{$\mu_{{\ell}m}=(\mu_{\ell}+\mu_{m})/2$},
where \mbox{$i,j,\ell,m=\overline{1,N}$}.

\end{document}